\documentclass[aps,pre,superscriptaddress]{revtex4-1}
\pdfoutput=1
\usepackage{graphicx}
\usepackage{amssymb}
\usepackage{amstext}
\usepackage{hyperref}
\usepackage{amsmath}
\renewcommand{\vec}[1]{\boldsymbol{\mathbf{#1}}}
\newcommand{\abs}[1]{| #1|}
\newcommand{\mat}[1]{\boldsymbol{\mathbf{#1}}}

\newcommand{\eqrefp}[1]{Eqs.\ \eqref{#1}}

\makeatletter
\renewcommand{\dot}[1]{{\mathop{\kern\z@#1}\limits^{\vbox to-1.4\ex@{\kern-\tw@\ex@\hbox{\normalfont .}\vss}}}}
\renewcommand{\ddot}[1]{{\mathop{\kern\z@#1}\limits^{\vbox to-1.4\ex@{\kern-\tw@\ex@\hbox{\normalfont ..}\vss}}}}
\renewcommand{\dddot}[1]{{\mathop{\kern\z@#1}\limits^{\vbox to-1.4\ex@{\kern-\tw@\ex@\hbox{\normalfont ...}\vss}}}}
\renewcommand{\ddddot}[1]{{\mathop{\kern\z@#1}\limits^{\vbox to-1.4\ex@{\kern-\tw@\ex@\hbox{\normalfont ....}\vss}}}}
\usepackage[capitalize]{cleveref}

\newcommand{\refcite}[1]{Ref.\ \cite{#1}}

\date{\today}
\hypersetup{
  pdfkeywords={},
  pdfsubject={},
  pdfcreator={Emacs 24.2.1 (Org mode 8.0.3)}}

\begin{document}
\title{Delay differential equation models for single and coupled bubble dynamics in a compressible liquid}
\author{Derek C. Thomas}
\email{dthomas@byu.edu}
\affiliation{Department of Physics and Astronomy, Brigham Young University, Provo, Utah 84602}
\affiliation{Applied Research Laboratories, The University of Texas at Austin, Austin, Texas 78713-8029}
\author{Yurii A. Ilinskii}
\affiliation{Applied Research Laboratories, The University of Texas at Austin, Austin, Texas 78713-8029}
\author{Mark F. Hamilton}
\affiliation{Applied Research Laboratories, The University of Texas at Austin, Austin, Texas 78713-8029}
\begin{abstract}
  Various models for interacting spherical bubbles in a compressible liquid based on delay differential equations are considered.  It is shown that most previously proposed models for interacting spherical bubbles in a compressible liquid based on the Keller-Miksis and Gilmore-Akulichev models are unstable for closely spaced bubbles.  A new model for a single spherical bubble in a compressible liquid is proposed and used to derive a stable model for interacting bubbles.  A qualitative comparison to the results of direct numerical integration of the fluid equations of motion suggests that the new model provides more accurate results than the standard Keller-Miksis or Gilmore-Akulichev models for single bubble dynamics.
\end{abstract}
\maketitle
\section{Introduction}
\label{sec:introduction}
A robust, stable time-domain model for coupled bubble motion in a compressible liquid is required for various applications.
Bubbles can significantly impact biomedical treatments including lithotripsy \cite{evan2002,bailey1998}, high intensity focused ultrasound \cite{bailey2001}, and histotripsy \cite{matlaga2008,xu2008}.
Treatment and diagnosis using ultrasound contrast agents also motivates the study of bubble dynamics.
Other applications include underwater and ocean acoustics,\cite{leighton2010,skaropoulos2003,kargl2002,feuillade1995} SONAR, ultrasonic cleaning \cite{niemczewski2007}, and sonochemistry \cite{suslick2003}.

Models for bubble dynamics consisting of ordinary differential equations for the different modes of oscillation (radial pulsation, translation, shape oscillations, etc.) have proven relatively successful in representing bubble dynamics \cite{lauterborn2010,prosperetti1986,plesset1977,lezzi1987,leightonTheAcousticBubble}.
We refer to these as discrete bubble models, that is, models in which the bubble is represented by a set of discrete, coupled, dynamical modes.
The canonical example of a discrete model for a single bubble is due to Rayleigh \cite{rayleigh1917} and Plesset \cite{plesset1949}.
This model has been modified and extended to account for liquid compressibility (radiation damping or energy lost to acoustic radiation)\cite{akulichev1968,gilmore1952,keller1956,miksis1989,leightonTheAcousticBubble,prosperetti1986,lezzi1987}, thermodynamic \cite{gilmore1952,devin1959,leightonTheAcousticBubble}, and viscous effects \cite{devin1959,ilinskii1992,ilinskii2007,doinikov2004,leightonTheAcousticBubble}.

Ordinary differential equation models for interacting bubbles in an incompressible liquid can be obtained by Lagrangian \cite{luther2000,doinikov2004,ilinskii2007} and Hamiltonian \cite{ilinskii2007} formulations.
Liquid compressibility is included in discrete bubble models as a delay in the bubble-bubble interaction terms.
Models for interacting bubbles in an compressible liquid have been derived primarily in the linear approximation as systems of coupled resonant scatterers \cite{tolstoy1986,tolstoy1988,feuillade1995,feuillade1996} or as {\it ad hoc} modifications of the incompressible case \cite{ilinskii1992,mettin2000,ilinskii2005}.
For frequency-domain models and eigenvalue analysis of linearized model equations, the delay in bubble interaction manifests as a complex phase coefficient.
For time-domain models, the system of ordinary differential equations is converted to a system of delay differential equations.

The majority of previous analyses have relied on the frequency domain approach or eigenvalue analyses. 
\citet{feuillade1995,feuillade1996} used a frequency domain analysis to show that liquid compressibility can have a significant impact on the damping of a bubble system, even for closely spaced bubbles.
In fact, it was shown that as the bubble separation distance approaches zero, the radiation damping also approaches zero for bubbles in antiphase motion.

It has been shown by eigenvalue analysis that the use of delay differential equation models can provide better agreement with experimental results \cite{ooi2008,doinikov2005,manasseh2004}.
However, studies based on the time domain integration of the delay differential equations of motion for a bubble system are relatively rare \cite{fujikawa1986,mettin2000,ilinskii2005}.
This may be due to the difficulties associated with integrating delay differential equations numerically.
Numerical integration of delay differential equations requires special tools \cite{bellen2003numerical}, and delay differential equations may exhibit unexpected behavior \cite{lakshmananDynamicsOfNonlinearTimeDelaySystems,erneuxAppliedDelayDifferentialEquations}.
Mathematical analysis of bubble models with time delay has shown that certain models exhibit instability \cite{heckman2009,sinden2011}.
However the causes and implications of these model instabilities has not been investigated.

The primary goals of this paper are first to illustrate instabilities in certain previously proposed models and second to present a new model based on a set of delay differential equations.
The new model appears to be more accurate for the single bubble case and does not possess the same unstable behavior for the single and interacting bubble cases.

\section{Motivation}
\label{sec:motivation}
We will show that most previously proposed models for the dynamics of coupled bubbles in a compressible liquid are unstable for time-domain simulation of closely spaced bubbles.
Because closely spaced bubbles do not exhibit unbounded growth, this instability is nonphysical and we view it as a deficiency of existing models that must be corrected.
System stability is required by energy conservation; in the absence of external forcing, each bubble begins with a finite amount of energy and cannot gain any energy.
In an incompressible, inviscid medium without thermal effects a bubble will oscillate indefinitely as the energy is converted between potential energy due to the compression of the gas in the bubble and kinetic energy due to the motion of the surrounding fluid.
No energy is radiated or lost from the system.
The same holds for a multibubble system, the total energy in the system is constant.
For a single bubble in a compressible medium, as the bubble oscillates it produces waves that carry energy away from the bubble; this produces a damping effect in the bubble motion, often called radiation damping \cite{devin1959,leightonTheAcousticBubble,medwinFundamentalsOfAcousticalOceanography,ilinskii1992}.

The inclusion of compressibility effects in models for the dynamics of systems of coupled bubbles requires a delay in bubble interaction to account for wave propagation.
It has been suggested that the ordinary differential equation (ODE) models for coupled bubble dynamics in an incompressible liquid can be used to obtain model equations for dynamics in a compressible liquid by incorporating the propagation delays \cite{fujikawa1986, ilinskii1992, ilinskii2005}.
This produces a delay differential equation (DDE) model for bubble dynamics in a compressible liquid.
A system of DDEs may have very different behavior from the related ODE system obtained by removing the delay, even becoming unstable \cite{erneuxAppliedDelayDifferentialEquations}.

Physical considerations dictate that a model for bubbles in a compressible liquid must remain stable.
The only difference between a bubble system in an incompressible liquid and the same system in a compressible liquid is the rate at which the system radiates (loses) energy.
In the incompressible liquid, the system does not radiate; in the compressible liquid it does.
This means that the peak oscillation amplitude of the system will decrease in a compressible liquid as energy is radiated.
This suggests a criterion of stability for valid bubble models.
 \citet{feuillade1995,feuillade1996} showed that the damping of a bubble system is very different with, and without, delayed interaction.
In fact, the radiation damping was shown to approach zero for equally-sized bubbles in antiphase motion.

The dominant mode of bubble motion is the radial pulsation mode, this is the only mode considered here.
Translation and higher-order shape oscillations are neglected.
Discrete bubble models are typically formulated in terms of the bubble radius $R$ or the bubble volume $V=4\pi R^3/3$.
For systems containing multiple bubbles, the variables associated with the current bubble are indexed by $i$ and the interactions with other bubbles in the system are represented by sums over the indices $j$ and $k$.
\subsection{Previously proposed models}
We begin with models formulated in terms of the radial displacement.
These models are generally of the following form for the nonlinear case \cite{ilinskii2005,hay2008}
\begin{align}
\label{eq:lagDampEOM}
\left( 1-\frac{\dot{R}_i}{c_0}  \right)R_i\ddot{R}_i+\frac{3}{2}\left( 1-\frac{\dot{R}_i}{3c_0} \right)\dot{R}_i^2&=\frac{1}{\rho_0}\left( 1+\frac{\dot{R}_i}{c_0} +\frac{R_i}{c_0}\frac{d}{dt}\right)\left( P_i-P_0-p_{ei} \right) \nonumber\\&\qquad- \sum_{i\neq j}\left[ \frac{R_j}{D_{ij}}\left( R_j\ddot{R}_j+2\dot{R}_j \right) \right]_{\tau_{ij}},
\end{align}
where $D_{ij}$ is the separation distance between the bubbles $i$ and $j$, $\tau_{ij}$ is the time required to propagate from bubble $j$ to bubble $i$ at the acoustic sound speed $c_0$, $P_0$ is the ambient pressure, $\rho_0$ is the liquid density, and the sum is over all the bubbles in the system subject to the indicated constraints.
The pressure inside the $i$th bubble is
\begin{equation}
P_i = P_0 \left( \frac{R_{0i}}{R_i} \right)^{ - 3   \gamma }.
\end{equation}
We employ brackets with a subscript to indicate delayed variables:
\begin{equation}
\label{eq:bracket-def}
[f]_{\tau} = f(t-\tau)
\end{equation}
Without the interaction terms \cref{eq:lagDampEOM} can be recognized as the Keller-Miksis equation for a single bubble in a compressible liquid \cite{keller1956,keller1980,prosperetti1986}.

The linearization of \cref{eq:lagDampEOM} is obtained by assuming that the bubble radius can be represented as $R_i(t) = R_{0i} + \xi_i(t)$ where $\xi_i$ is the radial displacement and retaining only terms that are linear in $\xi_i$.  The result is
\begin{equation} 
\label{eq:linPrevDampEOM}
\ddot{\xi}_i(t) + \omega_{0i}\delta_{i, \rm rad}\dot{\xi}_i(t) + \omega_{0i}^2\xi_i(t) =-\frac{p_{ei}(t)}{R_{0i}\rho_0} - \sum_{i\neq j}\frac{R_{0j}^2}{D_{ij}R_{0i}}\ddot{\xi}_j(t-\tau_{ij})
\end{equation}
where the Minnaert or natural oscillation frequency of a single bubble is given by $\omega_{0i}^2=3\gamma P_0/R_{0i}^2\rho_0$ where $\gamma$ is the polytropic constant or ratio of specific heats for the gas inside the bubble.
This model was used by  \citet{doinikov1997} and \citet{ooi2008}.

The instability in \cref{eq:linPrevDampEOM} is demonstrated by considering a system of two bubbles.
For a system of two bubbles of equal size separated by a distance $D$ without an external source, \cref{eq:linPrevDampEOM} produces a set of coupled equations:
\begin{subequations}
  \begin{align}
\ddot{\xi}_1(t)  + \omega_{0}\delta_{\rm rad}\dot{\xi}_1(t) + \omega_0^2\xi_1(t) + \frac{R_0}{D}\ddot{\xi}_2(t-\tau) &= 0\label{eq:twoBubLinEqA}\\
\ddot{\xi}_2(t)  + \omega_{0}\delta_{\rm rad}\dot{\xi}_2(t) + \omega_0^2\xi_1(t) + \frac{R_0}{D}\ddot{\xi}_1(t-\tau) &= 0\label{eq:twoBubLinEqB},
\end{align}
\end{subequations}
where the dimensionless radiation damping coefficient is $\delta_{\rm rad} = \omega_0R_0 / c_0$, and the interaction delay $\tau$ is given by
\begin{align}
\tau&=\frac{D}{c_0}.
\end{align}

\begin{figure}
\includegraphics[width=8cm]{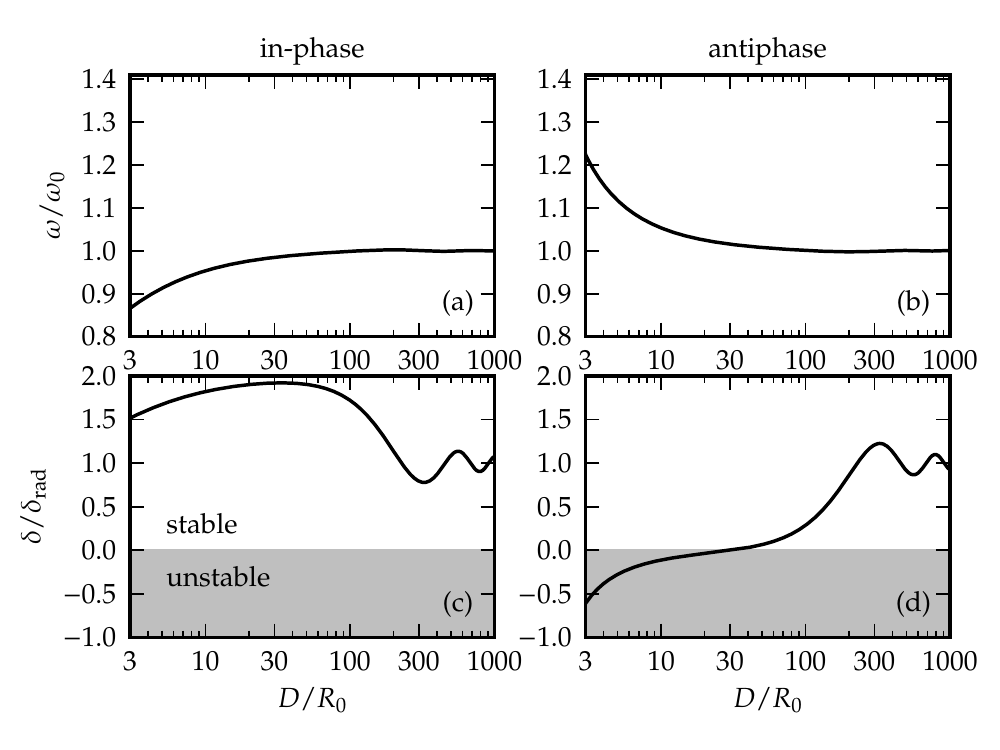}
\caption{\label{fig:two-bub-rad-damp-ana-prev}Comparison of normalized natural frequencies and damping coefficients as a function of the separation distance $D$ of the bubbles as predicted by \cref{eq:twoBubPrevApproxCharEqA,eq:twoBubPrevApproxCharEqB} for the two-bubble system described by \cref{eq:twoBubLinEqA,eq:twoBubLinEqB}.  The horizontal axis is the separation distance divided by the bubble radius.  Parts (a) and (b) show the natural frequency of the coupled system normalized by the natural frequency of a single bubble and parts (c) and (d) show the dimensionless damping coefficient for the in-phase and antiphase modes respectively.  Stable regions of the parameter space are indicated by a white background, unstable regions are indicated by a gray background.}
\end{figure}

A decoupled system of equations is obtained by adding and subtracting \cref{eq:twoBubLinEqA,eq:twoBubLinEqB} and defining $\xi_{+} = \xi_1 + \xi_2$ and $\xi_{-}=\xi_1-\xi_2$ \cite{feuillade1995,feuillade1996}.
The new variables correspond to the in-phase mode ($\xi_{ +}$) and the antiphase mode ($\xi_{-}$) of the system.
When written in terms of the new variables, the linearized equations of motion are
\begin{subequations}\label{eq:twoBubPrevApprox}
\begin{align}
\ddot{\xi}_+(t) + \omega_{0}\delta_{\rm rad}\dot{\xi}_+(t) + \omega_{0}^2\xi_+(t) + \frac{R_0}{D}\ddot{\xi}_+(t-\tau) &= 0,\\
\ddot{\xi}_-(t) + \omega_{0}\delta_{\rm rad}\dot{\xi}_-(t) + \omega_{0}^2\xi_+(t) - \frac{R_0}{D}\ddot{\xi}_-(t-\tau) &= 0.
\end{align}
\end{subequations}
The solutions to \cref{eq:twoBubPrevApprox} are assumed to be of the form $\xi=\xi_0 e^{\lambda t}$, where $\xi_0$ is a constant and $\lambda$ is an eigenvalue.
The characteristic equations for the in-phase and antiphase modes of \cref{eq:linPrevDampEOM} are
\begin{subequations}
\label{eq:twoBubPrevApproxCharEq}
\begin{align}
\lambda_+^2 \left( 1 + \frac{R_0}{D}e^{-\lambda_+\tau}\right) + \lambda_{+} \omega_0\delta_{\rm rad} + \omega_0^2 &= 0\label{eq:twoBubPrevApproxCharEqA},\\
\lambda_-^2 \left( 1 -\frac{R_0}{D}e^{-\lambda_-\tau}\right) + \lambda_{-} \omega_0\delta_{\rm rad} + \omega_0^2 &= 0  \label{eq:twoBubPrevApproxCharEqB}
  \end{align}
\end{subequations}
respectively.
These equations are transcendental equations with an infinite number of discrete eigenvalues and the equations must be solved numerically.
In general, the eigenvalues are complex, 
\begin{equation}
\label{eq:s-def}
\lambda=-\delta\omega/2 \pm i\omega,
\end{equation}
where $\delta$ is the dimensionless damping coefficient (reciprocal of the quality factor) and $\omega$ is the natural frequency of the corresponding mode.
\Cref{fig:two-bub-rad-damp-ana-prev} shows the numerically calculated natural frequency and damping coefficient of the first mode for the in-phase system (left) and the natural frequency and damping coefficient of the first unstable mode for the antiphase system.
The results in \cref{fig:two-bub-rad-damp-ana-prev} were obtained by applying a numerical root-finding algorithm to \eqrefp{eq:twoBubPrevApproxCharEq}.
It can be seen that antiphase motion of the system is unstable for closely spaced bubbles.
As stated previously, this instability represents a deficiency in this bubble model.

We now consider several previous models formulated in terms of the bubble volume.   \citet{ilinskii1992} proposed
\begin{align}
\label{eq:i-z-coupled}
\frac{\rho_0}{2(6\pi^2)^{2/3}}\left( \frac{\ddot{V}_i}{V_i^{1/3}} - \frac{\dot{V}_i^2}{V_i^{4/3}} \right) - \frac{\rho_0}{4\pi c_0}\dddot{V}_i = P_i - P_0 - \frac{\rho_0}{4\pi}\sum_{j\neq i}\frac{\ddot{V}_j(t-D_{ij}/c_0)}{D_{ij}}
\end{align}
as a volumetric model for bubble oscillation.
This model is unique because the radiation damping is represented by the $\rho_0\dddot{V}_i/4\pi c_0$ term.
 \Cref{eq:lagDampEOM} can be derived from \cref{eq:i-z-coupled} by converting from bubble volume to bubble radius and then iteratively differentiating and substituting while retaining terms to $O(c_0^{-1})$ and neglecting terms of order $O(c_0^{-1})\times O(R/D)$\cite{hay2008}.

\Cref{eq:i-z-coupled} can be linearized by assuming that the volume can be expressed as $V_i = V_{0i}+v_i$ and expanding all nonlinear terms to first order in the volume displacement $v_i$.  The result is
\begin{equation}
\label{eq:triple-feuillade}
\frac{\rho_0}{4\pi R_{0i}}\ddot{v}_i(t) - \frac{\rho_0}{4\pi R_{0i} c_0}\dddot{v}_i(t) + \frac{\rho_0\omega_{0i}^2}{4\pi R_{0i}} v_i(t)=-p_{ei}(t) - \sum_{i\neq j}\frac{\rho_0}{4\pi D_{ij}}\ddot{v}_j(t-\tau_{ij}).
\end{equation}
\Cref{eq:triple-feuillade} can be related to the model proposed by \citet{devin1959} for a single bubble and extended by \citet{feuillade2001,feuillade1995,feuillade1996} to include bubble interaction.
If the volume displacement is assumed to be time-harmonic ($v = v_0e^{i\omega t}$) then the second term in \cref{eq:triple-feuillade} can be rewritten as
\begin{equation}
 - \frac{\rho_0}{4\pi R_{0i} c_0}\dddot{v}_i(t) = \frac{\rho_0\omega^2}{4\pi R_{0i} c_0}\dot{v}_i(t).
\end{equation}
With this expression, \cref{eq:triple-feuillade} becomes
\begin{equation}
\label{eq:feuillade}
\frac{\rho_0}{4\pi R_{0i}}\ddot{v}_i(t) + b_i(\omega) \dot{v}_i(t) + \rho_0\frac{\omega_{0i}^2}{R_{0i}} v_i(t)=-p_{ei}(t) - \sum_{i\neq j}\frac{\rho_0}{4\pi D_{ij}}\ddot{v}_j(t-\tau_{ij})
\end{equation}
where $\tau_{ij} = D_{ij}/c_0$ and the damping coefficient $b_i(\omega)$ is given by $b_i(\omega)= \rho_0\omega^2/4\pi R_{0i}c_0$ (viscous and thermal damping are neglected).
\Cref{eq:feuillade} is the equation used by Feuillade; without the interaction terms it is the equation derived by Devin.
Because of the frequency dependent damping coefficient $b_i(\omega)$, the models proposed by Devin and Feuillade and shown in \cref{eq:feuillade} are only valid for time-harmonic motion.

It can be shown that \cref{eq:triple-feuillade} is unstable for time-domain integration by analyzing \cref{eq:triple-feuillade} for a single bubble
\begin{equation}
\label{eq:triple-d-feuillade-single}
\frac{\rho_0}{4\pi R_0}\ddot{v}(t) - \frac{\rho_0}{4\pi c_0}\dddot{v}(t) + \frac{\rho_0\omega_0^2}{4\pi R_0} v(t)=0.
\end{equation}
With the ansatz $v=v_0e^{\lambda t}$, where $v_0$ is a constant, the characteristic equation for the eigenvalue $\lambda$
\begin{equation}
\label{eq:tdfs-char}
 - \frac{1}{4\pi c_0}\lambda^3 + \frac{1}{4\pi R_0}\lambda^2 + \frac{\omega_0^2}{4\pi R_0}=0.
\end{equation}
can be obtained.

The roots of \cref{eq:tdfs-char} are
\begin{align}
\label{eq:tdfs-roots}
\lambda&=\frac{c_0}{3R_0} + \alpha \beta + \frac{\beta^{\ast} c_0^2}{9 R_0^2\alpha}
\end{align}
where 
\begin{equation}
\alpha=\left(\frac{c_0\omega_0}{R_0}\sqrt{\frac{\omega_0^2}{4}+\frac{c_0^2}{27R_0^2}} + \frac{c_0\omega_0^2}{2R_0} + \frac{c_0^3}{27R_0^3}
\right)^{\frac{1}{3}}.
\end{equation}
and $\beta^{\ast}$ is the complex conjugate of $\beta$; $\beta$ is chosen from
\begin{equation}
\label{eq:beta-def}
\beta=\left\{ -\frac{1}{2} \pm i \frac{\sqrt{3}}{2}, 1\right\}.
\end{equation}
The complex values of $\lambda$ correspond to oscillatory modes.
The natural frequency and damping coefficient can be calculated using \cref{eq:s-def}.
For a bubble with a radius of 10 $\mu$m, with $c_0=1482$ m/s, $P_0=101325$ Pa, $\rho_0=998$ kg/m$^3$, and $\gamma=1.4$, the damping coefficient is $\delta=0.0139$ which is precisely the value given by \citet{leightonTheAcousticBubble} for the dimensionless radiation damping coefficient of a single bubble.
The symbol $\delta_{\textrm{rad}}$ is used to represent the damping coefficient of a single bubble.
The real root of \cref{eq:tdfs-char} is positive and thus represents an unstable mode.
This instability will not be observed in purely time-harmonic systems; however, any transient excitation will excite this instability.
Unstable behavior is expected for third-order ordinary differential equations with a small leading coefficient.
The instability can be eliminated by iteratively differentiating and substituting while retaining terms to $O(c_0^{-1})$\cite{hay2008}.
The result can be shown to be equivalent to \cref{eq:lagDampEOM} and \cref{eq:linPrevDampEOM} for the nonlinear and linear cases, respectively and will thus be unstable for the coupled bubble problem.
A more extensive analysis of \cref{eq:linPrevDampEOM} and similar equations is presented in \refcite{thomas2012} along with series expansions that partially correct the instabilities.

It should be noted that unstable modes were not observed in previous work by  \citet{feuillade1995,feuillade1996,feuillade2001} and  \citet{ilinskii1992} because of the methods that were employed.
\citet{feuillade1995,feuillade1996,feuillade2001} assumed that the motion was time harmonic and that the oscillation frequency was given by the oscillation frequency of coupled bubbles in an incompressible liquid; fixing the oscillation frequency ensures a stable time-harmonic form.
The analysis shown \cite{ilinskii1992} also assumed a time-harmonic form and both interaction and nonlinear terms were included via expansion to a certain order.
The expansion employed produces unstable modes that are not oscillatory as was shown in \cref{eq:tdfs-char,eq:tdfs-roots}.
Therefore the time-harmonic assumption removes any unstable modes.

The work of  \citet{ooi2008} considered the transcendental nature of the characteristic equations but limited the search for eigenvalues to the neighborhood of the eigenvalues of the system in the incompressible limit; additionally, the damping due to viscous and thermal effects was sufficiently high to preclude the observation of unstable modes.
It appears that the conclusions presented in the work of Feuillade \cite{feuillade1995,feuillade1996,feuillade2001},  \citet{doinikov1997}, and  \citet{ooi2008} are valid for the parameter and frequency ranges that were considered.

In general, all modes of a system will be excited in a time-domain simulation and thus the unstable modes of these models will provide nonphysical results.
Additionally, even when the unstable modes are masked by damping they may affect the predicted dynamics.
Therefore, we consider alternative models for coupled bubble dynamics in a compressible liquid that are stable.

\subsection{Previous delay differential equation model for single-bubble dynamics}
\label{sec:prev-delay-diff}
\citet{ilinskii1992} obtained \cref{eq:i-z-coupled} from an approximation to a delay differential equation model that was derived from physical considerations.
This suggests that the instability in \cref{eq:i-z-coupled} is an artifact of the expansion used to obtain the approximate form; we will show that this is not the case.
 \citet{ilinskii1992} proposed the following model for a single bubble in a compressible liquid:
\begin{equation}
\label{eq:i-z-single}
\frac{\rho_0}{2(6\pi^2)^{2/3}}\left( \frac{[\ddot{V}]_\tau}{V^{1/3}} - \frac{\dot{V}^2}{V^{4/3}} \right) = P - P_0
\end{equation}
where $\tau=R/c_0$.
The derivation of this equation is motivated by the Li\'{e}nard-Weichert potential of classical electrodynamics \cite{jacksonClassicalElectrodynamics}.
Ilinskii and Zabolotskaya showed that this model correctly accounts for the energy lost to acoustic radiation.
\Cref{eq:i-z-coupled} is an approximation to this model that can be used to obtain the Keller-Miksis model as was discussed previously.

In order to analyze the stability of \cref{eq:i-z-single}, we seek a linearized form.
\Cref{eq:i-z-single} can be linearized by letting $V=V_0+v$ where $V_0=4\pi R_0^3/3$ to obtain
\begin{equation}
\label{eq:IandZlin}
\ddot{v}(t-R_{0}/c_0) +  \omega_{0}^2v(t)=0.
\end{equation}
The stability of \cref{eq:IandZlin} can be analyzed by assuming that the solution will be of the form $v=v_0e^{\lambda t}$.
The eigenvalue $\lambda$ is complex and is assumed to be of the form given in \cref{eq:s-def}.
The resulting characteristic equation is
\begin{equation}
\label{eq:IandZlinCharEq}
 \lambda^2 e^{-\lambda R_{0}/c_0} + \omega_0^2 = 0
\end{equation}
which can be solved analytically for the eigenvalues $\lambda$:
\begin{equation}
\label{eq:char_sol}
\lambda = -\frac{2 c_0}{R_0} W_n \left( - i R_0 \omega_0/2 c_0 \right),
\end{equation}
where $W_n(x)$ is the \(n\)th branch of the Lambert W or product log function.
The frequencies and damping coefficients corresponding to the real and imaginary parts of the eigenvalues $\lambda$ as given by \cref{eq:s-def} are shown in Table \ref{tab:IandZtab}.
Negative values of $\delta$ correspond to unstable modes.
It can be seen that the damping of the ``fundamental'' frequency (\(n=0\)) is equivalent to the standard dimensionless damping coefficient $\delta_{\rm rad} = 0.0139$ \cite{leightonTheAcousticBubble}.
The delayed self-action model for a single bubble given in \cref{eq:IandZlin} has unstable modes with very high frequencies ($\sim 768$ times the fundamental frequency).
In time domain simulations, arbitrary input can excite the unstable modes.
\begin{table}[htb!]
\caption{\label{tab:IandZtab}Natural frequency (relative to the Minnaert frequency for an undamped bubble) and dimensionless damping coefficient predicted by delayed self-action bubble model for modes corresponding to values of $n$ ranging from $-1$ to 1.  Modes with negative values of $\delta$ are unstable.}
\begin{tabular}{rrr}
$n$ & $\omega/\omega_{0}$ & $\delta$\\
\hline
-1 & 768.596079231 & -2.67312944025\\
0 & 0.99992762304 & 0.0139330079762\\
1 & -262.458684232 & 7.58759922187\\
\end{tabular}
\end{table}

\section{New model based on Hamiltonian formulation}
Motivated by the success of \cref{eq:i-z-single} in obtaining the correct asymptotic forms for a single bubble and correctly representing the energy lost to acoustic radiation, we follow a similar approach.
We begin with the Hamiltonian equations of motion for a system of coupled bubbles.
The Hamiltonian equations are given by  \citet{ilinskii2007} as
\begin{subequations}\label{eq:eom}
\begin{align}
\Dot{R}_i&=\frac{1}{4\pi\rho_0}\left[ \frac{G_i}{R_i^3} - \sum\limits_{j\neq i}\frac{G_j}{R_iR_jD_{ij}} + \sum\limits_{k\neq i,j}\frac{R_kG_j}{R_iR_jD_{ik}D_{jk}} \right],\label{eq:eomR}\\
  \Dot{G}_i&=\frac{1}{4\pi\rho_0} \left[ \frac{3}{2}\frac{G_i^2}{R_i^4} - \sum\limits_{j\neq i} \frac{G_iG_j}{R_i^2R_jD_{ij}} + \sum\limits_{k\neq i,j} \frac{R_kG_iG_j}{R_i^2R_jD_{ik}D_{jk}} \right.
\nonumber
\\
&\qquad\qquad-\left. \frac{1}{2}\sum\limits_{i\neq j,k}\frac{G_iG_k}{R_jR_kD_{ij}D_{ik}} \right]
\nonumber
\\
&
\qquad+4\pi R_i^2 \left(P_i - P_0 -p_{ei} \right).\label{eq:eomG}
\end{align}
\end{subequations}
Here $G_i$ is used to represent the radial momentum of a bubble, or in other words, the momentum conjugate to the radial state variable.
It should be noted that there is a mistake in the last term of the second equation in \refcite{ilinskii2007}; the restrictions on the ranges of the indices are given there as $k\neq i, j$, whereas they should be $i\neq j, k$.

\subsection{Single bubble model}
It can be shown that the delayed term in \cref{eq:i-z-single} is identical to the interaction terms in \cref{eq:i-z-coupled} with the separation distance $D=R_i$.
In other words, the method proposed by \citet{ilinskii1992} implies that the effect of liquid compressibility can by included by introducing a delayed self-action term of the same form as the standard interaction terms.
We apply this reasoning to the Hamiltonian equations of motion and introduce delayed self-action terms that have the same form as the regular interaction terms.
The equations are modified so that in the incompressible limit, the original equations are recovered.
The result of this procedure is
\begin{subequations}\label{eq:ham-delay-single}
\begin{align}
\dot{R}_i &= \frac{1}{4\pi\rho_0}\left[ \frac{G_i}{R_i^3}- \frac{[G_i]_{\tau_i}}{R_i [R_i]_{\tau_i}^2}+ \frac{G_i}{R_i^2   [R_i]_{\tau_i}} \right]\label{eq:ham-delay-rad}\\
\dot{G}_i &= \frac{1}{4\pi\rho_0}\left[ 2 \frac{G_i^2}{R_i^4} - \frac{[G_i]_{\tau_i} G_i} { R_i^3[R_i]_{\tau_i}}+ \frac{G_i^2}{[R_i]_{\tau_i}   R_i^3}- \frac{1}{2}   \frac{[G_i]_{\tau_i} G_i}{[R_i]_{\tau_i}^2   R_i^2} \right]\nonumber\\
&\qquad+ 4\pi R_i^2  ( P_i - P_0 - p_{ei})\label{eq:ham-delay-rad-mom}
\end{align}
\end{subequations}
where 
\begin{align}
\tau_i &= R_i/ c_0.
\end{align}

\begin{figure}[htb]
\centering
\includegraphics[width=8cm]{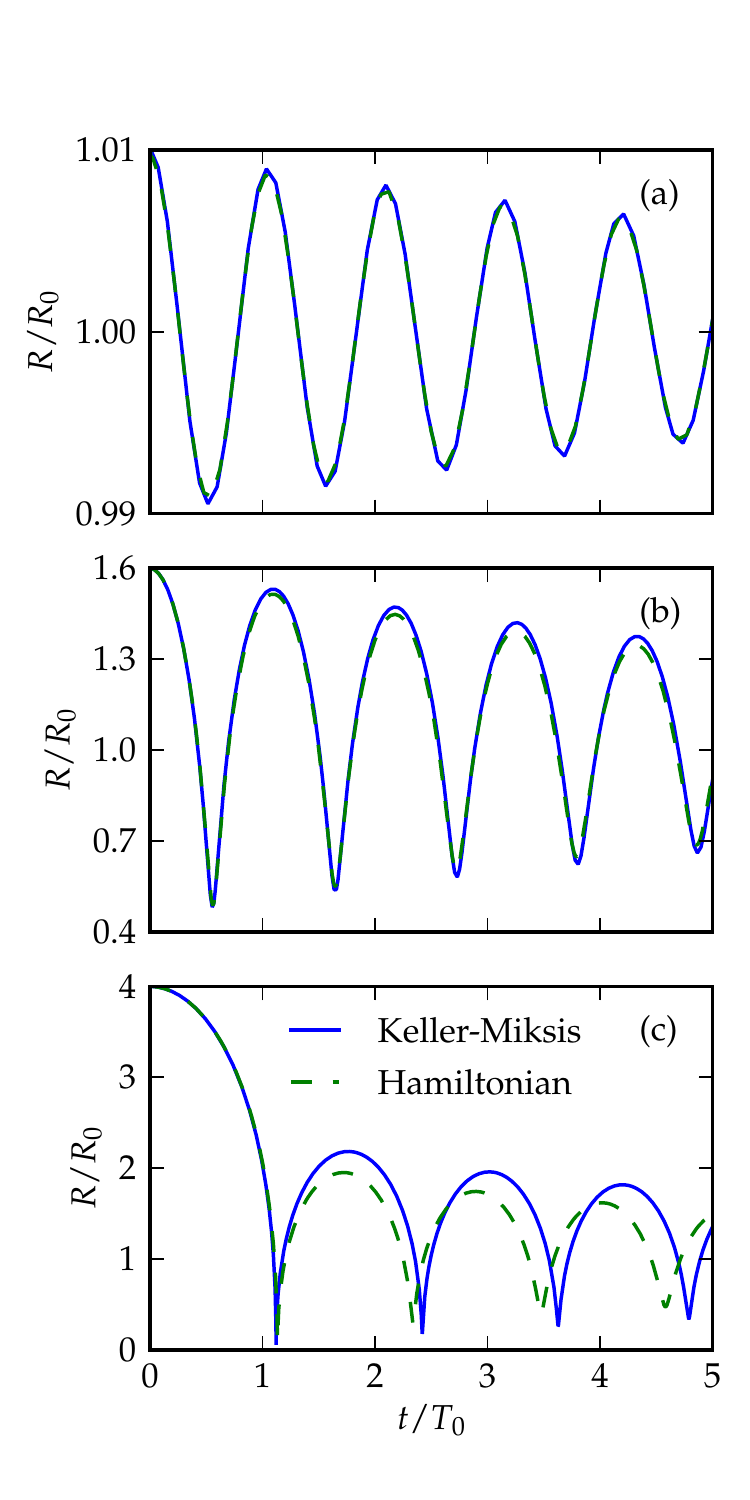}
\caption{\label{fig:km-compare}Comparison of new model with Keller-Miksis model for three different initial conditions: $R=1.01R_0$, $R=1.7R_0$, and $R = 4R_0$.}
\end{figure}

\Cref{fig:km-compare} shows the results of numerical integration of the Hamiltonian model proposed here and the Keller-Miksis model for a single bubble in free response.
Viscosity and surface tension are neglected.
All numerical integration is carried out using the RADAR5 package \cite{guglielmi2001}.
Three initial conditions are shown, $R=1.01R_0$, $R=1.7R_0$, and $R = 4R_0$.
It can be seen that for low amplitude, $R=1.01R_0$ (part a), the two models agree to within graphical precision.
For moderate amplitude, $R=1.7R_0$ (part b), the models agree well but there are minor differences.
These differences are likely due to differences at higher order in $c_0^{-1}$ (the Keller-Miksis model is only valid to first order).
The differences between the predictions of the two models become much more significant at high amplitudes, $R = 4R_0$ (part c).

Comparison with previous results presented by \cite{fuster2011a} suggests that the new model is more accurate.
\citet{fuster2011a} considered various models for a single bubble in a compressible medium and compared them to a complete model based on direct integration of the Navier-Stokes equations for bubbles undergoing violent collapse.
Unfortunately, insufficient information about the models and the relevant physical constants used in the paper was included to reproduce the results for direct  comparison to the model derived here.
Instead we make a qualitative comparison to the results presented there.
The initial conditions in the fourth figure of \citet{fuster2011a} are such that the Keller-Miksis model predicts a rebound of 54\% of the original radius; in contrast, the complete fluid model predicts a rebound to an amplitude of 44\% of the original radius (the other models, Gilmore-Akulichev, Tomita-Shima, have similar rebounds).
\Cref{fig:km-compare} part c shows the response predicted for initial conditions for which the Keller-Miksis model produces a rebound with an amplitude of approximately 55\% of the initial radius.
With the same initial conditions the delayed Hamiltonian model presented here predicts an initial rebound to approximately 46\% of the initial radius, which is much closer to the prediction of the complete model; this suggests that the new model is more accurate. .
As presented here, the delayed Hamiltonian model does not appear to rely on any expansions in $c_0^{-1}$.
This may improve the accuracy of predictions during rebound.

It is possible to derive an approximate form of \cref{eq:ham-delay-rad,eq:ham-delay-rad-mom} by assuming that $\tau_i$ is much smaller than the time scales of interest.
A Taylor expansion of \cref{eq:ham-delay-rad,eq:ham-delay-rad-mom} about $\tau_i=0$ yields
\begin{subequations}\label{eq:ham-delay-single-approx}
\begin{align}
\Dot{R}_i&=\frac{1}{4\pi\rho_0}\left[ \frac{G_i}{R_i^3} + \frac{1}{c_0}\left( \frac{\dot{G}_i}{R_i^2} - \frac{\dot{R}_iG_i}{R_i^3} \right)\right]\label{eq:ham-delay-single-approx-rad}\\
\dot{G}_i &= \frac{1}{4\pi\rho_0}\left[ \frac{3}{2} \frac{G_i^2}{R_i^4} - \frac{1}{c_0}\left( \frac{G_i^2\dot{R}_i}{R_i^4} - \frac{3}{2} \frac{G_i\dot{G}_i}{R_i^3}\right) \right]\nonumber\\
&\qquad+ 4\pi R_i^2  ( P_i - P_0 - p_{ei})\label{eq:ham-delay-single-approx-mom}.
\end{align}
\end{subequations}
It can be seen that in the incompressible limit, $c_0\rightarrow \infty$, the Hamiltonian equations for a single bubble in an incompressible liquid as given by \cite{ilinskii2007} are recovered
\begin{subequations}\label{eq:ham-incomp-single}
\begin{align}
\Dot{R}_i&=\frac{1}{4\pi\rho_0} \frac{G_i}{R_i^3}\label{eq:ham-incomp-single-rad}\\
\dot{G}_i &= \frac{1}{4\pi\rho_0} \frac{3}{2} \frac{G_i^2}{R_i^4}+ 4\pi R_i^2  ( P_i - P_0 - p_{ei})\label{eq:ham-incomp-single-mom}.
\end{align}
\end{subequations}
The approximate expressions in \eqrefp{eq:ham-delay-single-approx} can be written in a form suitable for integration by standard methods for ODEs by defining the matrix
\begin{equation}
\label{eq:mat-ham-delay-single-approx}
\mat{M} =
\begin{bmatrix}
  1&0\\
0&1
\end{bmatrix}
+ \frac{1}{4\pi R_i^4 \rho_0c_0}
\begin{bmatrix}
  R_iG_i & - R_i^2\\
G_i^2 & -\frac{3}{2}R_iG_i
\end{bmatrix}.
\end{equation}
With this matrix, \eqrefp{eq:ham-delay-single-approx} can be written in vector form as
\begin{equation}
\label{eq:ham-delay-single-approx-vec-form}
\begin{bmatrix}
  \dot{R}_i\\
\dot{G}_i
\end{bmatrix}
= \mat{M}^{-1}
\begin{bmatrix}
  \textrm{RHS(\cref{eq:ham-incomp-single-rad})}\\
  \textrm{RHS(\cref{eq:ham-incomp-single-mom})}
\end{bmatrix}.
\end{equation}
Numerical integration has shown that this form provides accurate results for all cases tested.
\subsection{Bubble interaction}
\begin{figure}[htb]
\centering
\includegraphics[width=8cm]{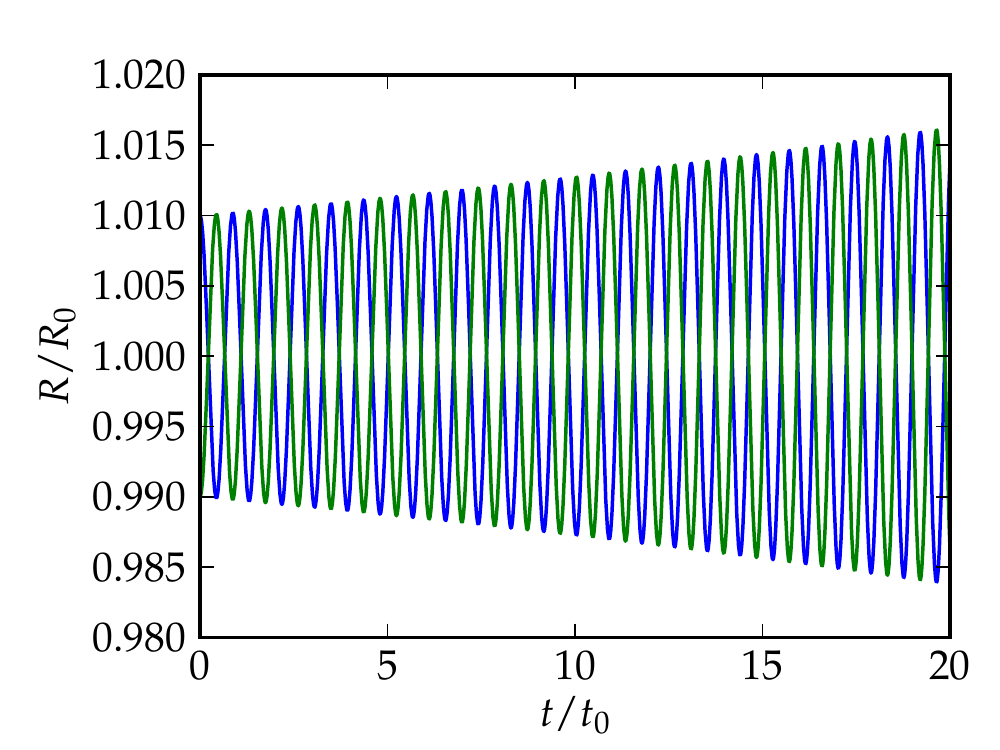}
\caption{\label{fig:two-bub-unstable}Bubble radii for a system of two bubbles of equal size in antiphase motion.  The bubbles are separated by a distance of $3 R_0$.  Clearly the model is unstable.}
\end{figure}

We now discuss how the effects of bubble interaction are incorporated into \cref{eq:ham-delay-rad-mom,eq:ham-delay-rad} to create a model for interacting bubbles.
The most obvious approach is to add terms 2 and 3 in \cref{eq:eomR}) to \cref{eq:ham-delay-rad} for the radius and terms 2, 3, and 4 in \cref{eq:eomG} to \cref{eq:ham-delay-rad-mom} for the radial momentum with the appropriate delay in the interaction terms.
This approach is discussed extensively in \cite{thomas2012}.
The results of numerical integration shown in \cref{fig:two-bub-unstable} demonstrate that this approach produces an unstable model and so we present an alternative.

Note that it is clear where the effect of additional acoustic pressure sources should appear in \cref{eq:ham-delay-rad,eq:ham-delay-rad-mom}.
The pressure produced by the motion of a spherical surface of radius $R$ centered at the origin is \cite[p. 155]{pierceAcoustics}
\begin{align}\label{eq:pt-src-press}
p(\vec{r},t) &= \frac{\rho_0}{4\pi \abs{\vec{r}}}\ddot{V}(t-\abs{\vec{r}} / c_0)
\end{align}
where $V$ is the volume and
\begin{equation}\label{eq:ddV}
\ddot{V} = 4\pi \left( R^2 \ddot{R} + 2R \dot{R}^2\right).
\end{equation}
It is not always appreciated that \cref{eq:pt-src-press} is not restricted to infinitesimal oscillations \cite{strasberg1956}.
\Cref{eq:pt-src-press,eq:ddV} together provide an expression for the pressure produced by the motion of a bubble.
The pressure due to other bubbles in the system acting on the $i$th bubble is inserted into the bubble equations of motion as an additional pressure source.
Only the radial momentum equation contains pressure terms and so only the radial momentum equation is modified to account for bubble interactions.
With the interaction pressure given by \cref{eq:pt-src-press,eq:ddV}, the equation for the radial momentum becomes
\begin{align}\label{eq:ham-delay-rad-mom-coupled}
\dot{G}_i &= \frac{1}{4\pi\rho_0}\left[ 2 \frac{G_i^2}{R_i^4} 
- \frac{ G_i [G_i]_{\tau_i}} {R_i [R_i]_{\tau_i}^3}
+ \frac{ G_i^2} {R_i^3 [R_i]_{\tau_i}}
- \frac{1}{2}   \frac{[G_i]_{\tau_i} G_i}{[R_i]_{\tau_i}^2   R_i^2} \right]\nonumber\\
&\qquad+ 4\pi R_i^2  \left[ P_i - P_0 - p_{ei} - \sum_{j\neq i}\frac{\rho_0}{D_{ij}}\left( [R_j]_{\tau_{ij}}^2 [\ddot{R}_j]_{\tau_{ij}} + 2[R_j]_{\tau_{ij}} [\dot{R}_j]_{\tau_{ij}}^2\right)\right]
\end{align}
where
\begin{align}
\tau_{ij} &= D_{ij} / c_0.
\end{align}

The presence of the delayed second derivative of the bubble radius $[\ddot{R}_j]_{\tau_{ij}}$ on the right-hand side of \cref{eq:ham-delay-rad-mom-coupled} prevents numerical integration of \cref{eq:ham-delay-rad,eq:ham-delay-rad-mom-coupled} by standard methods.
An alternate expression for $\ddot{R}_j$ in terms of $R_i$, $G_i$, $\dot{R}_i$, and $\dot{G}_i$ can be obtained by differentiating \cref{eq:ham-delay-rad} with respect to time, letting $i=j$, and then delaying the result by $\tau_{ij}$ to obtain

\begin{align}
[\ddot{R}_j]_{\tau_{ij}} &= 
\frac{1}{4\pi\rho_0}\left[
\frac{[\dot{G}_j]_{\tau_{ij}}}{[R_j^3]_{\tau_{ij}}}
- 3 \frac{[\dot{R}_j]_{\tau_{ij}}[G_j]_{\tau_{ij}}}{[R_j^4]_{\tau_{ij}}}
\right.\nonumber\\
&\qquad\qquad
+2 \frac{[\dot{R}_j]_{\sigma_{ij}}[G_j]_{\sigma_{ij}}}{[R_j]_{\tau_{ij}}[R_j]_{\sigma_{ij}}^3}
-2 \frac{[\dot{R}_j]_{\tau_{ij}}[G_j]_{\tau_{ij}}}{[R_j]_{\tau_{ij}}^3[R_j]_{\sigma_{ij}}}
\nonumber\\
&\qquad\qquad
+\frac{[\dot{R}_j]_{\tau_{ij}}[G_j]_{\sigma_{ij}}}{[R_j]_{\tau_{ij}}^2[R_j]_{\sigma_{ij}}^2}
-\frac{[\dot{R}_j]_{\sigma_{ij}}[G_j]_{\tau_{ij}}}{[R_j]_{\tau_{ij}}^2[R_j]_{\sigma_{ij}}^2}
\nonumber\\
&\qquad\qquad
+\frac{[\dot{G}_j]_{\tau_{ij}}}{[R_j]_{\tau_{ij}}^2[R_j]_{\tau_{i}}}
-\frac{[\dot{G}_j]_{\sigma_{ij}}}{[R_j]_{\tau_{ij}}[R_j]_{\sigma_{ij}}^2}
\nonumber\\
&\qquad\qquad
+\frac{1}{c_0}\left( 
\frac{[\dot{R}_j]_{\tau_{ij}}[\dot{G}_j]_{\sigma_{ij}}}{[R_j]_{\tau_{ij}}[R_j]_{\sigma_{ij}}^2}
+ \frac{[\dot{R}_j]_{\tau_{ij}}[\dot{R}_j]_{\sigma_{ij}}[G_j]_{\tau_{ij}}}{[R_j]_{\tau_{ij}}^2[R_j]_{\sigma_{ij}}^2}
\right.\nonumber\\
&\qquad\qquad\qquad\quad
\left.\left.
-2 \frac{[\dot{R}_j]_{\tau_{ij}}[\dot{R}_j]_{\sigma_{ij}}[G_j]_{\sigma_{ij}}}{[R_j]_{\tau_{ij}}[R_j]_{\sigma_{ij}}^3}
 \right)
\right];\label{eq:ddrj-delay}
\end{align}

Note the presence of the double delay
\begin{align}
\sigma_{ij} &= \tau_{ij} + [R_j]_{\tau_{ij}} / c_0\nonumber\\
&=(D_{ij} + [R_j]_{\tau_{ij}}) / c_0
\end{align}
in \cref{eq:ddrj-delay}.
With the expression for $[\ddot{R}_j]_{\tau_{ij}}$ given in \cref{eq:ddrj-delay}, \cref{eq:ham-delay-rad,eq:ham-delay-rad-mom-coupled} together define a system of coupled, neutral, delay differential equations with state-dependent delays that describes the dynamics of coupled bubbles.

In the incompressible limit, \cref{eq:ham-incomp-single-rad} gives the relationship between the bubble radius and the radial momentum which can be solved to obtain an expression for the radial momentum $G_i = 4\pi \rho_0 R_i^3 \dot{R}_i$.
With this expression for $G_i$ and without the effect of the external source, it can be shown that \cref{eq:ham-delay-rad-mom-coupled} reduces to the modified Rayleigh-Plesset equation for coupled bubbles in an incompressible liquid given by  \citet{ilinskii2007}
\begin{align}
\label{eq:lag-incomp}
R_i\ddot{R}_i + \frac{3}{2}\dot{R}_i^2&= \frac{P_i-P_0}{\rho_0} - \sum_{j\neq i}\frac{R_j}{D_{ij}}\left( R_j\ddot{R}_j + 2 \dot{R}_j\right).
\end{align}
Thus we see that the standard bubble model is recovered in the incompressible limit \cite{ilinskii2007}.

In order to linearize \cref{eq:ham-delay-rad-mom-coupled,eq:ham-delay-rad} let
\begin{align}
R_i &= R_{0i} + \xi_i,\\
G_i &= \eta_i.
\end{align}
The linearized form of \cref{eq:ham-delay-rad} is
\begin{equation}
\label{eq:ham-delay-rad-lin}
\dot{\xi}_i = \frac{1}{4\pi\rho_0 R_{0i}^3}\left( 2\eta_i - [\eta_i]_{\tilde{\tau}_i} \right)
\end{equation}
where $\tilde{\tau}_i = R_{0i} / c_0$.
The linearized form of \cref{eq:ham-delay-rad-mom-coupled} is
\begin{equation}
\label{eq:ham-delay-rad-mom-coupled-lin}
\dot{\eta}_i = -4\pi R_{0i}^2 \left[ 3\gamma P_0 \xi_i + p_{ei} + \rho_0R_{0i}^2\sum_{j\neq i}\frac{[\ddot{\xi}_j]_{\tilde{\tau}_{ij}}}{D_{ij}}\right]
\end{equation}
where $\tilde{\tau}_{ij} = D_{ij} / c_0$.
A single equation can be obtained by differentiating \cref{eq:ham-delay-rad-lin} with respect to $t$ and substituting \cref{eq:ham-delay-rad-mom-coupled-lin} on the right-hand side:
\begin{align}
\label{eq:ham-delay-lin}
\ddot{\xi}_i &=  \omega_{0i}^2 \left([\xi_i]_{\tilde{\tau}_i}-2\xi_i\right) + \frac{1}{\rho_0R_{0i}} \left( [p_{ei}]_{\tilde{\tau}_i} - 2 p_{ei} \right) \nonumber\\
&\qquad\qquad+ \sum_{j\neq i}\frac{[\ddot{\xi}_j]_{\tilde{\tau}_{ij}+\tilde{\tau}_j}-2[\ddot{\xi}_j]_{\tilde{\tau}_{ij}}}{R_{0i}D_{ij}}.
\end{align}
If the external acoustic pressure $p_{ei}$ is neglected, the characteristic equation for \cref{eq:ham-delay-lin}is
\begin{equation}
\label{eq:ham-delay-char}
\lambda^2 \left( 1 - R_{0i}\sum_{j\neq i}\frac{\xi_{0j}}{\xi_{0i}}\frac{e^{-\lambda(\tilde{\tau}_{ij} + \tilde{\tau}_j)} - 2e^{-\lambda\tilde{\tau}_{ij}}}{D_{ij}} \right) - \frac{3\gamma P_0}{\rho_0R_{0i}} \left( e^{-\lambda\tilde{\tau}_i} - 2 \right) = 0.
\end{equation}
\Cref{eq:ham-delay-char} is a transcendental equation that cannot be solved analytically.
The roots of this equation may lie far in the right half of the complex plane and thus be difficult to find numerically.
We instead rely on numerical integration of the equations of motion to test for stability.

\begin{figure}[htb]
\centering
\includegraphics[width=8cm]{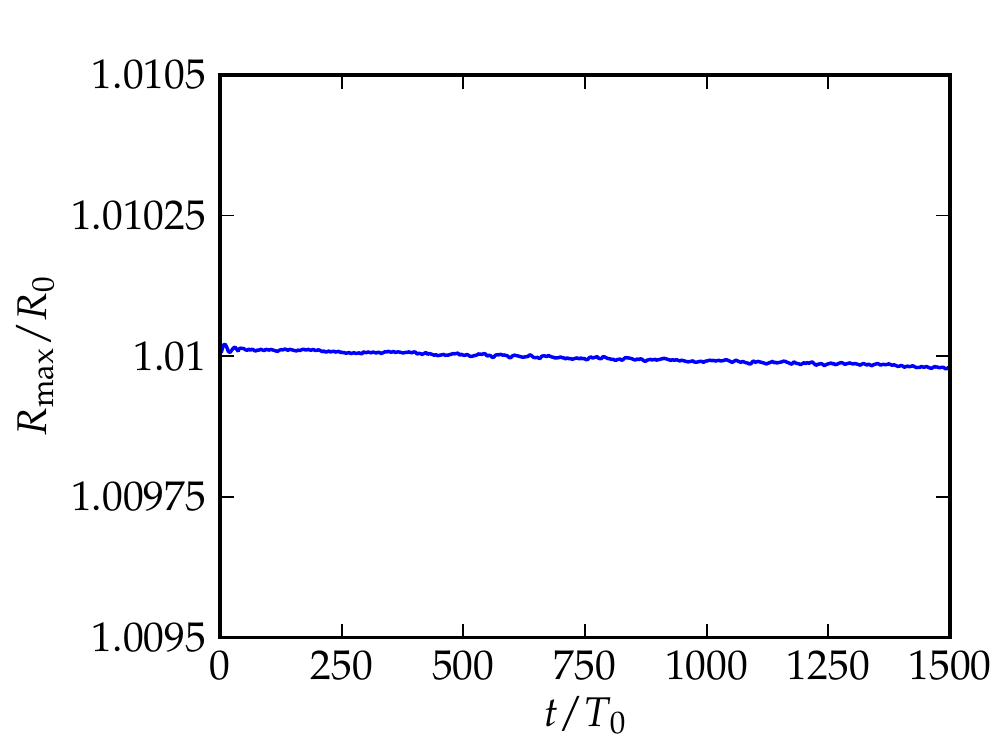}
\caption{\label{fig:two-bub-stable-max}Maximum radius per period for a system of two bubbles of equal size in antiphase motion.  The bubbles are separated by a distance of $3 R_0$.  The gradual decrease in the radial amplitude indicates that the system is stable, although the damping is very low.  The slight variations in the line are due to numerical sampling.}
\end{figure}

Unstable modes manifest themselves in the time-domain solution; therefore rather than performing a numerical search of the complex plane for unstable eigenvalues, we simply integrate the equations of motion over a sufficiently long time interval to detect unstable modes.
The most significant test of stability is the case of two bubbles of equal size in antiphase motion.
This is the case that was shown to be unstable for the standard bubble model in \cref{fig:two-bub-rad-damp-ana-prev}.
The results from the integration for a system of two bubbles in antiphase motion are shown in \cref{fig:two-bub-stable-max}.
The system consists of two bubbles with an equilibrium radius of $10\mu$m separated by a distance of $30\mu$m ($3 R_0$).
One bubble radius is initially at $1.01R_0$ while the other is at $0.99 R_0$.
Because the integration is carried out over such a long time, direct analysis of the bubble radii as a function of time is not particularly illuminating.
Instead the maximum bubble radius for each period is shown.
Clearly the model is stable, even for closely spaced bubbles.
This agrees well with the conclusions of  \citet{feuillade1996,feuillade1995}, who showed that the radiation damping of a pair of equally sized bubbles in antiphase motion approaches zero.
Although it was shown previously that Feuillade's model is unstable for general time-domain integration, his model is valid for systems in perfect time-harmonic motion without transient disturbances.

\section{Conclusions}
\label{sec:conclusions}
We have shown that several previous models for the dynamics of interacting spherical bubbles are unstable.
This instability prevents numerical integration of the model equations in the time domain for closely spaced bubbles.
We have shown that the model for single bubble dynamics with delayed self-action due to liquid compressibility proposed by \citet{ilinskii1992} is unstable although it can be used to obtain correct asymptotic forms.
A new single bubble model based on delayed self-action has been developed.
The new model is stable and agrees with the predictions of the Keller-Miksis model for low amplitude motion.
As the amplitude increases, the models begin to diverge.
\citet{fuster2011a} showed that the Keller-Miksis, Gilmore-Akulichev, and Tomita-Shima models underpredict the damping due to liquid compressibility by comparing to the results of numerical integration of the fluid equations of motion.
Our new model predicts more damping for high amplitude motion than the other discrete bubble models and is qualitatively closer to the predictions of the direct numerical simulation, thus suggesting that the new model better represents the physical system.

We have also derived a model for coupled bubbles in a compressible liquid based on the new single bubble model.
This model relies only on the assumption that the bubbles are spherical and the assumption that disturbances in the host liquid propagate at the equilibrium sound speed $c_0$.
The new model for coupled bubbles is stable where the previously proposed models are not.
Thus a stable time domain model for coupled bubble dynamics in a compressible liquid with arbitrary initial conditions and input has been obtained.
We believe that the new delay differential equation models for bubble dynamics presented here provide useful alternatives to the Keller-Miksis model and similar discrete models for bubble dynamics in a compressible liquid.

%

\end{document}